\documentclass{iopjournal}
\usepackage[utf8]{inputenc}
\usepackage{booktabs}
\usepackage{graphicx}
\usepackage{dcolumn}
\usepackage{bm}
\usepackage[utf8]{inputenc} 
\usepackage{amsmath}
\usepackage{lmodern}
\begin{document}

\articletype{Paper} 

\title{Photons in Media: A Second-Quantization Scheme Based on the Optical Dirac Equation}

\author{Lili Yang$^1$, Longlong Feng$^1$ and Pengming Zhang$^1$}

\affil{$^1$School of Physics and Astronomy, Sun Yat-Sen University, Zhuhai 519082, China}

\email{yanglli7@mail.sysu.edu.cn, flonglong@mail.sysu.edu.cn, zhangpm5@mail.sysu.edu.cn}


\begin{abstract}

We develop a second-quantization framework for photons based on optical Dirac equation of source-free Maxwell theory in generic media, in which the electromagnetic field is recast as a four-component spinor-like wave function admitting both positive- and negative-energy solutions, which are naturally interpreted as photon and anti-photon states. Expanding the field in terms of single-photon eigenmodes, we construct a fully consistent quantization scheme in which the photon field operators obey bosonic commutation relations, in close analogy with the Dirac quantization of electrons. In structured media, the optical Dirac equation acquires effective mass and coupling terms induced by the dielectric tensor, analogue to an electronic-Dirac-type structure. This allows the propagation of photons in media to be interpreted in terms of boosted spinor states, providing a unified description of vacuum and medium-modified dispersion relations. The framework further reveals a natural quantum-mechanical origin of transverse spin in structured electromagnetic fields, including evanescent waves, where spin components perpendicular to the propagation direction emerge from the underlying helicity structure. In context of optical Dirac theory,  this work presents a quantum field theoretic description of photons in both vacuum and media, offering a new perspective on photon quantization, spin–orbit interaction, and light–matter coupling in structured optical systems.

\end{abstract}

%
%
%
%
%
%
%

\section{Introduction}
Maxwell’s equations can be recast in Schrödinger- or Dirac-like form by introducing the Riemann--Silberstein complex field vectors constructed from the electric field $\mathbf{E}$ and the magnetic field $\mathbf{B}$. Such reformulations naturally admit an interpretation in terms of a photon wave function \cite{Darwin1932NotesOT,BialynickiBirula1994OnTW,BIALYNICKIBIRULA1996245,Laporte1931ApplicationOS,Moses1959SolutionOM,cohentannoudji1989PhotonsAA,Barnett_2014,PhysRevLett.120.243605,BIALYNICKIBIRULA2006342,KELLER20051,Horsley2018TopologyAT,Cugnon2011ThePW,Elbistan2016DualityAH,Yamamoto_2017}, thereby providing a bridge between the classical field description of light and its quantum-mechanical properties. Within this wave-function representation, key observables of the electromagnetic field, including energy, momentum, total angular momentum, and spin, arise naturally as expectation values of the corresponding operators acting on the photon wave function \cite{cohentannoudji1989PhotonsAA,KELLER20051,Enk1994SpinAO,Kobe1999ARS,Bliokh2013MagnetoelectricEI,Bliokh2012DualEH,Bliokh_2015}. This framework therefore makes it possible to discuss these quantities from a single-photon perspective.

In our previous work, we formulated the electromagnetic field in structured media as a wave function, such that Maxwell’s equations can be cast into a Dirac-like form \cite{PhysRevA.106.043513,yang2025inducedberryconnectionphotonic}. Based on this formulation, we develop a second-quantization scheme for the electromagnetic field in spin-degenerate media using the corresponding optical Dirac-like equation. The electromagnetic field is expanded in terms of single-photon eigenstates obtained as solutions of Maxwell’s equations \cite{Adlard1997LOCALIZATIONOO,BialynickiBirula1998ExponentialLO}, which provide a natural basis for field quantization. The corresponding photon wave function is further expanded in the effective spin basis, where the expansion coefficients are identified with photon and antiphoton creation and annihilation operators obeying bosonic commutation relations. Consequently, the standard classical expressions for the energy, momentum, angular momentum, and spin of the electromagnetic field are recovered consistently in their second-quantized representations. This construction also provides a natural quantum-mechanical interpretation of the transverse spin degree of freedom in structured fields, such as evanescent waves \cite{Bliokh:2013xla}. Moreover, the formal correspondence between photon energies in vacuum and in media, together with the analogy to free and accelerated electrons, suggests that a photon effectively acquires a “boost” upon entering a medium, reflecting the modification of its dispersion relation. Extending this Dirac-like framework, photon–medium interaction can be viewed as analogous to the dynamics of electrons in an external magnetic field. In this picture, the photonic wave function plays the role of the electronic wave function, while the electromagnetic vector potential is replaced by an effective gauge field induced by the medium. This correspondence is particularly relevant for structured photonic systems such as helical waveguides, where the spin–orbit interaction of light can be interpreted as a coupling between the photonic wave function and this dual gauge field \cite{yang2025spinorbitcouplinghelicalwaveguides}. The formulation naturally admits negative-energy solutions and provides a unified description of photon dynamics in media. For convenience, the natural unit system is adopted throughout this paper.

\section{Quantization of Free Photons}
The positive and negative photon wave functions $\psi^{\pm}$ also referred to as canonical variables, are defined as follows \cite{KELLER20051}:
\begin{equation}
\begin{aligned}\label{canonical variables}
& \psi^+(\mathbf{r},t)=\mathbf{E}_{\perp} (\mathbf{r},t)+ \mathbf{e}_{\mathbf{k}} \times \mathbf{B}(\mathbf{r},t),\\
& \psi^- (\mathbf{r},t)=\mathbf{E}_{\perp} (\mathbf{r},t)- \mathbf{e}_{\mathbf{k}} \times \mathbf{B}(\mathbf{r},t).
\end{aligned}
\end{equation}
Here, $ \bm e_{\mathbf{k}}$ denotes the propagation direction, while the subscript $\bot$ indicates the transverse field. These canonical variables (representing photons and anti-photons) satisfy the following dynamical equations:
\begin{equation}
    \begin{aligned}\label{antiphotons}
        i\hbar \frac{\partial }{\partial t} \psi^+ = \hat{H }\cdot \psi^+, \quad i\hbar \frac{\partial }{\partial t} \psi^- = \hat{H }_A\cdot \psi^-,
    \end{aligned}
\end{equation}
where $\hat{H}$ is the Hamiltonian operator, with corresponding photon energy eigenvalue $E = \hbar\omega$, while the anti-photon energy is $E_A = -E$\cite{KELLER20051}. In this canonical-variable representation, physical quantities of the electromagnetic field can also be expressed as expectation values of the corresponding wave-function operators \cite{KELLER20051}. 
We can always choose a set of orthogonal basis vectors $\mathbf{e}_{\perp}$ perpendicular to the propagation direction $\mathbf{e}_{\mathbf{k}}$. For transverse fields, the electromagnetic field lies entirely within the plane spanned by $\mathbf{e}_{\perp}$ and can thus be decomposed in this basis. Denoting the basis vectors as $(\mathbf{e}_{x_1}, \mathbf{e}_{x_2})$, they can be further transformed into the helical representation, $ \mathbf{e}_{\pm} = (\mathbf{e}_{x_1} \pm i \mathbf{e}_{x_2})/\sqrt{2}$. These vectors satisfy the following helicity-basis property:
\begin{equation}\label{helicity eigenstates}
    (\mathbf{e}_{\mathbf{k}}\cdot \hat{\mathbf{s}})\mathbf{e}_{\pm}= \pm \mathbf{e}_{\pm},
\end{equation}
where the spin operator $\hat{\mathbf{s}}$ is represented in the adjoint form, $(\hat{s}_i)_{jk} = -i\epsilon_{ijk}$. By means of a suitable unitary transformation, the aforesaid helicity basis vectors can be projected onto the helicity eigenspace and take the explicit two-dimensional representation
\begin{equation}
    \begin{aligned}
 \mathbf{e}_+ = \binom{1}{0},\quad \mathbf{e}_-=\binom{0}{1}.
    \end{aligned}
\end{equation} 
In this representation, the projection of the spin operator $\hat{\bm s}$ onto the propagation unit vector $\mathbf{e}_{\mathbf{k}}$ reduces exactly to the Pauli matrix $\sigma_3$. For convenience, the propagation direction is often chosen to align with the $z$-axis. In this case, the transverse basis vectors are simply $\mathbf{e}_{\perp} = \{\mathbf{e}_x, \mathbf{e}_y\}$. Accordingly, the spatial components of the canonical variables introduced in Eq.~\eqref{canonical variables} take the following form in the helicity eigenspace:
\begin{equation}
    \begin{aligned}\label{wavefunction}
       \psi^+_{\perp}(\mathbf{ r}) =  E_{\perp}(\mathbf{ r}) + i\sigma_3 B_{\perp}(\mathbf{ r}),\\ 
      \psi^-_{\perp}(\mathbf{ r}) =E_{\perp}(\mathbf{ r}) -i\sigma_3 B_{\perp}(\mathbf{ r}).
    \end{aligned}
\end{equation}
Here we adopt a separation ansatz for the temporal and spatial dependencies of the field wavefunction, namely $\psi(\bm r,t)\propto \psi(\bm r )e^{-i\omega t} $.

By introducing a four-component wave function $\psi =(\psi^+,\psi^-)^{T}$, We obtain a optical-Dirac equation from Maxwell's equations under the paraxial approximation in vacuum\cite{PhysRevA.106.043513}: 
\begin{equation}
    \begin{aligned}\label{hamidun}
        i\frac{\psi(r,t)}{\partial t} = (\beta \hat {m} +\beta \mathbf{\alpha} \cdot \mathbf{\hat{p}}) \psi(r,t), 
    \end{aligned}
\end{equation}
where $\beta = \gamma^0$, $\boldsymbol{\alpha} = \gamma^0 \boldsymbol{\gamma}$, and
\begin{equation}
    \begin{aligned}
       & \hat{m}= \hat{k}_z + \frac{1}{2k_z} \hat{\mathbf{k}}^2_{\perp} , \\ 
       & \hat{p}_{\pm}= \frac{1}{k_z} k^2_{\pm},\quad \hat{p}_z=0 .
    \end{aligned}
\end{equation}
Here, the momentum operator reads $\hat{k}_i=-i\partial_i$, and the transverse momentum operators in the helical basis are defined as  $\hat{k}_{\pm}=(\hat{k}_x\mp \hat{k}_y)/\sqrt{2}$, with $\mathbf{\hat{k}}^2_{\perp}=\hat{k}^2_x + k^2_y$. The operator $\hat{m}$ serves as an effective mass. For a plane wave with vanishing transverse momentum, the energy relation reduces to $\omega = k_z$, analogous to the energy–mass equivalence $E=m$ in electronic systems. In contrast, the momentum operator $\hat{\mathbf{p}}$ couples positive and negative energy states, giving rise to the Zitterbewegung effect\cite{Unal1997,KOBE19997,10.1119/1.12819,PhysRevA.80.032118,PhysRevA.105.062211}.


A magnetic field $\bm B$ is always a transverse field, whereas an electric field $\bm E$ is not necessarily so. For example, in the presence of external charges, a longitudinal component of the electric field can exist. Under the Coulomb gauge, the longitudinal field $E_{\|}$ which is parallel to the propagation direction, is determined by the Coulomb potential. This longitudinal component contributes to physical quantities such as the energy and momentum of the electromagnetic field \cite{cohen}. However, as the present study focuses primarily on the intrinsic properties of the electromagnetic field, it is assumed that no external charges are present. Consequently, only the contributions from the transverse components of the field are considered. When extending the discussion to a medium, however, the effects of bound charges must also be taken into account. Therefore, in the presence of a medium, the displacement field $\bm D$ is used instead of the electric field $\bm E $ to ensure the proper transverse structure of the wave function.


In the leading order of $k_z$, the photon wave function can be expanded in terms of plane waves, similar to the treatment of the electronic wave function\cite{Peskin1995AnIT,Williams2022IntroductionTQ}:
\begin{equation}\label{bohanshu zhankai} \psi(r, t) = \int d^3k , 1/N(k) \sum_s \left( a_s(k) \mathbf{e}_s(k) e^{-i\omega t + i\mathbf{k} \cdot \mathbf{r}} + b^\dagger_s(k) \widetilde{\mathbf{e}}_s(k) e^{i\omega t - i\mathbf{k} \cdot \mathbf{r}} \right), \end{equation}
with normalization factor: 
\begin{equation}
    \begin{aligned}
        N(k)=\sqrt{2E_k(2\pi)^3}, \quad E_k =\omega.
    \end{aligned}
\end{equation}
The operators $a^{\dagger}_s, b^{\dagger}_s$ represent creation operators for photons and anti-photons, respectively, with $s = \pm $ denoting right- and left-handed helicities. Due to the bosonic nature of photons, these operators obey commutation relations (in contrast to the fermionic anti-commutation relations for electrons):
\begin{equation}
    \begin{aligned}\label{commutation relation}
        [a_s(\mathbf{k}),a^{\dagger}_{s^{\prime}}(\mathbf{k}^{\prime})]= (2\pi)^3\delta_{ss^{\prime}}\delta (\mathbf{k}-\mathbf{k}^{\prime}) = [b_s(\mathbf{k}),b^{\dagger}_{s^{\prime}}(\mathbf{k}^{\prime}) ].
    \end{aligned}
\end{equation}
Similar to the case of free electrons, we can introduce four sets of basis vectors to describe the spin states of photons and anti-photons. The four-component spin basis vectors for photons are defined as
\begin{equation}\label{jishiliang +} \mathbf{e}_+ = \left( \begin{array}{c} 1 \ 0 \ 0 \ 0 \end{array} \right)^T, \quad \mathbf{e}_- = \left( \begin{array}{c} 0 \ 1 \ 0 \ 0 \end{array} \right)^T. \end{equation}
while those for anti-photons are given by
\begin{equation}\label{jishiliang -} \mathbf{\widetilde{e}}_+ = \left( \begin{array}{c} 0 \ 0 \ 1 \ 0 \end{array} \right)^T, \quad \mathbf{\widetilde{e}}_- = \left( \begin{array}{c} 0 \ 0 \ 0 \ 1 \end{array} \right)^T. \end{equation}
These basis vectors satisfy the orthonormality relations
\begin{equation}\label{orthonormality e_s}
\mathbf{e}_s \cdot \mathbf{e}^*_{s^{\prime}} = \delta_{ss^{\prime}}, \quad \mathbf{\widetilde{e}}_s \cdot \mathbf{\widetilde{e}}^*_{s^{\prime}} = \delta_{ss'}, \quad \mathbf{\widetilde{e}}_s \cdot \mathbf{e}^*_{s^{\prime}} = \mathbf{e}_s \cdot \mathbf{\widetilde{e}}^*_{s^{\prime}} = 0. 
\end{equation}
Here, the Kronecker delta $\delta_{ss^{\prime}}$ enforces orthogonality between states with the same spin but equal energy, while states corresponding to different energies remain mutually orthogonal.

We introduce momentum-dependent basis vectors in Eq.~\eqref{bohanshu zhankai}, analogous to the spinor solutions $u(p)$ and $v(p)$ in electronic systems. Their relation to the fundamental basis vectors $\mathbf{e}_s$ and $\mathbf{\widetilde{e}}_s$ is expressed as\cite{Williams2022IntroductionTQ}
\begin{equation}
    \mathbf{e}_s(k_z) = A \mathbf{e}_s , \quad \widetilde{\mathbf{e}}_s(k_z) = A \widetilde{\mathbf{e}}_s,
\end{equation}
where $A$ is normalization constant. To be consistent with the spinor formalism in Peskin and Schroeder's book\cite{Peskin1995AnIT}, we choose $A=\sqrt{2\lambda k_z }$ such that their orthogonality relations satisfy
\begin{equation}
    \mathbf{e}_s^{\dagger}(k) \mathbf{e}_{s^{\prime}}(k) = 2E \delta_{ss^{\prime}} =2\lambda k_z\delta_{ss^{\prime}}.
\end{equation}

\section{Energy, Momentum, Spin, and Orbital Angular Momentum}\label{momentum}
In general, the photon wavefunction can be formulated within the Riemann–Silberstein (R–S) representation. It has been established that the quantum-mechanical expectation values of energy, momentum, spin, and orbital angular momentum coincide with their classical counterparts derived from the electromagnetic field.Since the electric field $\mathbf{E}(\mathbf{r})$ and magnetic field $\mathbf{B}(\mathbf{r})$ are real observables, they can be expressed in terms of the R–S wavefunction $\psi$ and its complex conjugate. The equivalence between the classical field quantities and their wavefunction representations is proven in Appendix \ref{app3}, where they are reformulated in terms of creation and annihilation operators.In the following, we evaluate the expectation values of these operators by introducing single-photon states. 

In the presence of free charges, the longitudinal component of the electric field, $E_{\|}$ contributes to the energy $H_{long}$ and momentum $P_{long}$ of the electromagnetic field, being directly related to the external charges—for instance, the energy of the longitudinal field corresponds to the Coulomb potential \cite{cohen}. However, this work focuses primarily on the intrinsic properties of the electromagnetic field. In the absence of external charges, only the contributions from the transverse components of the field need to be considered, and the subscript is henceforth omitted.
The Hamiltonian of the system is defined as
\begin{equation} \label{section: Hamiltonian}H = \int d^3r \psi^\dagger i\gamma^0 \frac{\partial \psi}{\partial t}. \end{equation}
It can be seen that this expression differs from the Hamiltonian of an electronic system by the presence of a $\gamma^0$ factor (see Table~\ref{table}). Substituting the mode expansion of the wavefunction $\psi(\mathbf{r},t)$ given in Eq.~\eqref{bohanshu zhankai}, the Hamiltonian can be expressed in terms of the creation and annihilation operators:
\begin{equation}
    H =\int d^3k \sum_s E_k (a^{\dagger}_s a_s + b^{\dagger}_sb_s) + (constant).
\end{equation}
In this derivation, we have employed the commutation relations given in Eq.~\eqref{commutation relation} together with the orthonormality condition in Eq.~\eqref{orthonormality e_s}, which enables the Hamiltonian to be expressed in a normal-ordered form. The resulting structure is directly analogous to the Hamiltonian of a simple harmonic oscillator.

Since the photon is its own antiparticle, an antiphoton can be interpreted as a photon traveling in the opposite direction. The photon states are defined in positive momentum space as $ \vert \mathbf{k},s \rangle = a^{\dagger}_s(\mathbf{k}) \vert 0 \rangle$  and in negative momentum space as $\vert -\mathbf{k},s \rangle = b^{\dagger}_s(\mathbf{k}) \vert 0 \rangle$. These states satisfy the following orthonormality condition:
\begin{equation}
    \begin{aligned}
        \langle \mathbf{k}, s \vert \mathbf{k}^{\prime} , s^{\prime } \rangle = \delta^{3} \mathbf{(k-k^{\prime })} \delta_{ss^{\prime}}.
    \end{aligned}
\end{equation}
The one-photon state is defined as\cite{BialynickiBirula1998ExponentialLO}:
\begin{equation}
    \begin{aligned}
        \vert 1ph \rangle = \int d^3k \left(f_s(\mathbf{k} ) a^{\dagger}_s(\mathbf{k}) \vert 0 \rangle + f^{\prime}_s(\mathbf{k} ) b^{\dagger}_s(\mathbf{k}) \vert 0 \rangle\right).
    \end{aligned}
\end{equation}
The state is normalized such that:
\begin{equation}
    \begin{aligned}
        \langle 1ph \vert 1ph \rangle = \int d^3k \left(f^*_s(\mathbf{k}) f_s(\mathbf{k} ) + (f^{\prime}_s(\mathbf{k} ))^* f^{\prime}_s(\mathbf{k})\right)=1.
    \end{aligned}
\end{equation}
The energy expectation value for the one-photon state is:
\begin{equation}
    \begin{aligned}
        \langle 1ph \vert H \vert 1ph \rangle = \int d^3 p \sum_s E_p(f_sf^*_s + f^{\prime}_s (f^{\prime}_s)^*) .
    \end{aligned}
\end{equation} 
For monochromatic light with frequency $\omega$, the expectation value of the Hamiltonian in a single-photon state is
\begin{equation}
    \begin{aligned}
        \langle 1ph \vert H \vert 1ph \rangle = \hbar \omega. 
    \end{aligned}
\end{equation} 

The momentum operator is
\begin{equation}
\begin{aligned}\label{QM:momentum}
        \mathbf{P} =  \int d^3 r \psi^{*}(\mathbf{r}) \gamma^0( -i \boldsymbol{ \nabla} )  \psi(\mathbf{r}) = \int d^3 k \sum_s\,\mathbf{k}\, (a^{\dagger}_s a_s + b^{\dagger}_s b_s),
    \end{aligned}
\end{equation}
where the factor $\gamma^0$ arises because our Hamiltonian is $\gamma^0$-pseudo-Hermitian. Specifically, the Hamiltonian $H$ in Eq.~(\ref{hamidun}) satisfies a special Hermitian-conjugation relation known as the $\gamma^0$-pseudo-Hermitian condition\cite{BENDER2005333,Ashida2020NonHermitianP}:
\begin{equation}\label{gamma0}
    \gamma^0  H \gamma^0=H^{\dagger}.
\end{equation}
It is straightforward to verify that when only diagonal elements are present, an operator $\hat{m}$ satisfies the Hermitian property in this sense, i.e.,$\hat{m}^\dagger = \gamma^0 \hat{m} \gamma^0.$ Furthermore, using the anti-commutation relations of the gamma matrices, $\gamma_+ = -\gamma_-$ and $\gamma_- = -\gamma_+$, one finds $\gamma^0 p_\perp \gamma_\perp \gamma^0 = \big(p_\perp \gamma_\perp\big)^\dagger.$ Hence, Eq.~\eqref{gamma0} is satisfied. In this system, $\gamma^0$ modifies the inner product as
\begin{equation}
    \langle \psi_1, \psi_2 \rangle^{\gamma^0} = \langle \psi_1, \gamma^0 \psi_2 \rangle.
\end{equation}
For a single-photon state $\vert 1\text{ph} \rangle$, the expectation value of the momentum reads
\begin{equation}
\begin{aligned}\label{QM:exp,momentum}
        \langle 1ph \vert \mathbf{P} \vert 1ph\rangle =\int d^3 k \,\sum_s\mathbf{k}\, \left( f^*_s f_s + (f^{\prime}_s)^* f^{\prime}_s \right).
    \end{aligned}
\end{equation}

The orbital angular momentum  $\mathbf{L}$ can be expressed in terms of the photon wavefunction and the creation and annihilation operators as
\begin{equation}
    \begin{aligned}
        \mathbf{L} &=\int d^3 r \,\psi^* \gamma^0 [\mathbf{r}\times (-i\boldsymbol{\nabla})]  \psi \\ 
        &= \int d^3 k \sum_s (\mathbf{r}\times \mathbf{ k})(a^{\dagger}_s a_s+b^{\dagger}_s b_s ).
    \end{aligned}   
\end{equation}

Similarly, the expectation value of the orbital angular momentum for a single-photon state $\vert 1ph \rangle$ is
\begin{equation}
    \langle 1ph\vert \hat{ \bm L } \vert 1ph\rangle= \int d^3 k \sum_s (\mathbf{r}\times \mathbf{k} )\left( f^*_s f_s + (f^{\prime}_s)^* f^{\prime}_s \right),
\end{equation}
which is consistent with the classical relation
\begin{equation}
    \mathbf{L} =\langle \mathbf{r} \rangle \times \bf{P}.
\end{equation}
where $ \langle \mathbf{r} \rangle$ is the expectation value of the position and $\mathbf{P}$ is the momentum in (\ref{QM:exp,momentum})

The transverse components of the electromagnetic field contribute exclusively to the spin aligned with the propagation direction. The spin operator, expressed in terms of the creation and annihilation operators of photons and antiphotons (see Appendix~\ref{app3}), is given by
\begin{equation}
    \begin{aligned}\label{spin operator}
        \mathbf{S}= \int d^3 k \mathbf{\hat{k}} (a_+^{\dagger} a_+ - a_-^{\dagger}a_- - b_+^{\dagger}b_+ + b_-^{\dagger}b_-) ,
    \end{aligned}
\end{equation}
where $\mathbf{\hat{k}} = \mathbf{k}/|\mathbf{k}|$ denotes the unit vector along the momentum direction. 

A one-photon state with helicity $s$ is defined as
\begin{equation}
    \vert 1ph \rangle_s = \int d^3 k (f_s(\mathbf{k}) a^{\dagger}_s \vert 0 \rangle + f^{\prime}_{-s}(\mathbf{k}) b^{\dagger}_{-s}) \vert 0 \rangle,
\end{equation}
and the expectation value of the spin operator in this state is
\begin{equation}
    \begin{aligned}
       \langle 1ph\vert \mathbf{S} \vert 1ph \rangle_s = \int d^3p\, s \mathbf{\hat{k}}\, \left(f_s^*f_{s}+(f^{\prime}_{-s})^*f^{\prime}_{-s}\right).
    \end{aligned}
\end{equation}
This result indicates that a positive-helicity state consists of both positive-helicity photons and negative-helicity antiphotons.

\section{The Spinor-Like Basis for Photons in a Medium}
In the following, we extend the discussion to the wavefunction formalism in a medium and briefly compare it with the quantization procedure of the Dirac theory.
In a medium, the photon wavefunction is more appropriately defined by replacing $\mathbf{E}$ with $\mathbf{D}$, since the transverse condition is automatically satisfied in the absence of free charges. In this scenario, positive and negative photon states are generally coupled. For simplicity, we consider a spin-degenerate medium with a constant dielectric tensor, where the spin-degeneracy condition is encoded in the tensor as\cite{PhysRevA.106.043513}
\begin{equation}
    \begin{aligned}
       \mu^{-1} = \epsilon^{-1}=  \left(\begin{array}{ccc}
Q_0 & Q_{+2} & Q_{+1} \\
Q_{-2} & Q_0 & Q_{-1} \\
Q_{-1} & Q_{+1} & q_0
\end{array}\right).
    \end{aligned}
\end{equation}
where the decomposition leads to two monopole moments $Q_0=\frac{1}{2}(\epsilon^{-1}_{11}+\epsilon^{-1}_{22})$ and $q_0=\epsilon^{-1}_{33}$, dipole $Q_{\pm 1}=\frac{1}{\sqrt{2}}(\epsilon^{-1}_{13}\mp i\epsilon^{-1}_{23})$ that form a dipole vector ${\bf q}=\{Q_{+1},Q_{-1}\}$, and quadrupole $Q_{ \pm 2}=\frac{1}{2}\left[\left(\epsilon^{-1}_{11}-\epsilon^{-1}_{22}\right) \mp i\epsilon^{-1}_{12}\right]$, corresponding to the scalar (spin-0), vector (spin-1), and tensor (spin-2) modes, respectively\cite{PhysRevA.106.043513}. 

Within this medium, the optical Dirac equation reads\cite{PhysRevA.106.043513}:
\begin{equation}
    \begin{aligned}
        i\frac{\partial\psi(r,t)}{\partial t} = (\beta \hat {m} +\beta \boldsymbol{\alpha} \cdot \mathbf{\hat{p}}) \psi(r,t), 
    \end{aligned}
\end{equation}
where $\beta=\gamma^0$ and $\boldsymbol{\alpha} =\gamma^0\boldsymbol{\gamma}$ are defined in the Dirac representation. The effective-mass operator reads
\begin{equation}
    \begin{aligned}\label{mass}
        \hat{m}=Q_0 \hat{k}_z+\frac{q_0}{ 2k_z} \hat{\mathbf{k}}_{\perp}^2-\mathbf{q} \cdot \hat{\mathbf{k}}_{\perp},
    \end{aligned}
\end{equation}
while the momentum operators are expressed as
\begin{equation}
    \begin{aligned}
      \hat{p}_{ \pm}=Q_{ \pm 2} \hat{k}_z- 2 Q_{ \pm 1} \hat{k}_{ \pm}+\frac{q_0}{k_z} \hat{k}_{ \pm}^2, \quad \hat{p}_z=0 . 
    \end{aligned}
\end{equation}
For the positive-energy solutions, we assume the relation $\omega = \lambda k_z$. The corresponding basis vectors are given by
\begin{equation}
    \begin{aligned}\label{Dirac:orthogonal relation} 
    \mathbf{e}_+(k) = \frac{k_z}{\sqrt{(\lambda + Q_0)k_z} }\left(\begin{array}{l}
        \lambda + Q_0  \\
       \quad 0\\ 
        \quad 0\\ 
        - Q_{-2}    
    \end{array}\right), \quad \mathbf{e}_-(k) = \frac{k_z}{\sqrt{(\lambda + Q_0)k_z} }\left(\begin{array}{l}
        \quad 0 \\
        \lambda + Q_0   \\ 
        - Q_{+2} \\ 
        \quad 0    
    \end{array} \right).
    \end{aligned}
\end{equation}
The leading-order dispersion relation in $k_z$ can be expressed in a form that closely parallels the energy–momentum relation for electrons:
\begin{equation}
    \begin{aligned}
        &\lambda^2 k^2_z = Q^2_0k^2_z - \vert Q_2\vert^2k^2_z,  \\ 
        &p^2_0 = m^2 + p^2.
    \end{aligned}
\end{equation}
Here, $E_p=p_0$ represents the energy of the electron, $m$ is the mass, and $p$ is the momentum. It can be concluded by comparison that $\lambda $ is a parameter analogous to the energy in electronic systems, $Q_2$ is analogous to the momentum, and  $Q_0 $  corresponds to the mass in the case of an electron. Of course, the eigenstates discussed above can also be expressed in spherical coordinates. In this representation, the order parameters are parameterized as $\lambda = Q_0 \cos \theta $ ,$Q_{\pm 2 } = Q_0 \sin \theta e^{\pm i\phi}$\cite{PhysRevA.106.043513}.

\begin{figure}[htbp]
            \centering
            \includegraphics[height=0.28\textheight]{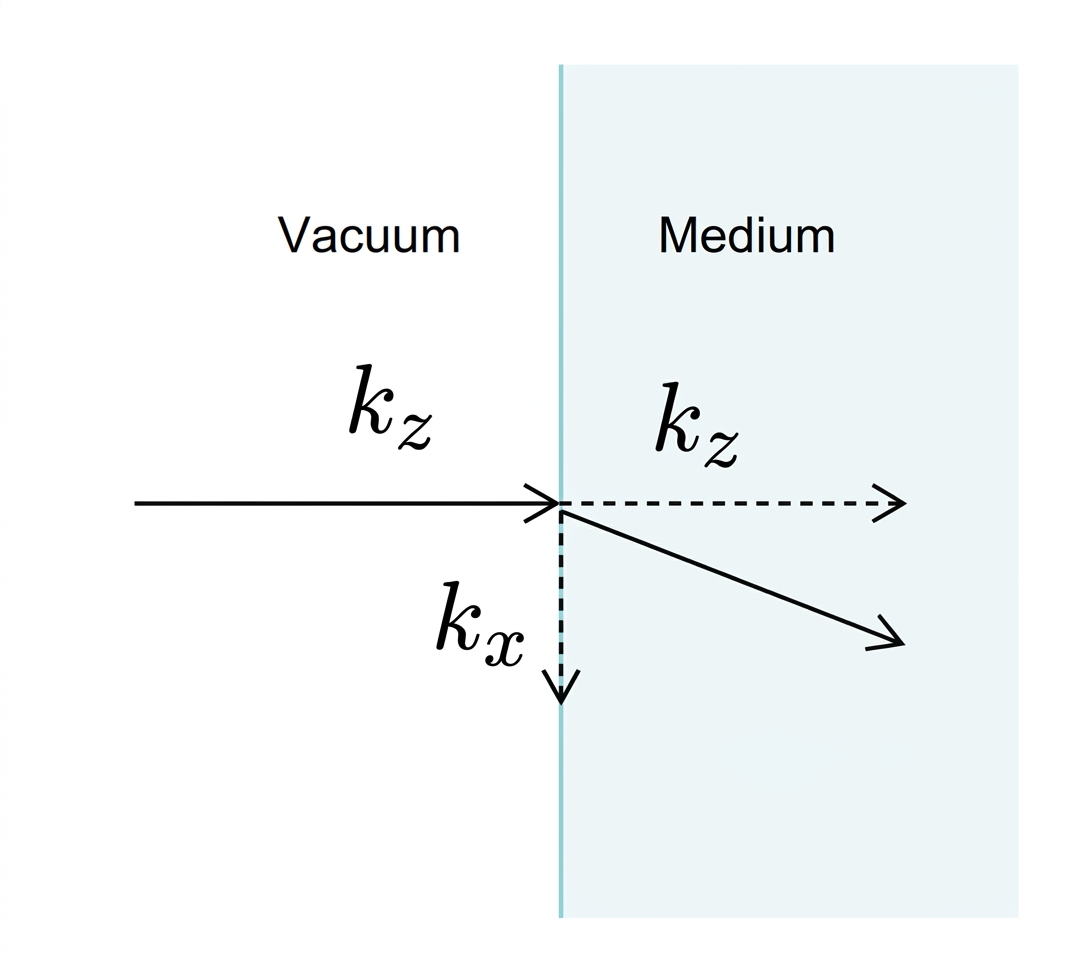}
             \caption{As shown in the figure, when free light propagates from left to right and enters the medium, the photon's trajectory deviates due to the inhomogeneity of the medium, which is equivalent to acquiring an effective transverse momentum.}
             \label{fig:boost}
\end{figure}

This formulation differs from conventional electronic systems: accelerating an electron increases its total energy above its rest-mass energy. In contrast, photon energy decreases when it acquires effective momentum, implying an effective mass and energy transfer to the medium. Similar to the Lorentz boost in electronic systems, which transforms the spinor basis of a stationary electron into that corresponding to an arbitrary momentum state, an analogous "boost" operation can be defined for photonic systems. This operation enables a photon to acquire a finite transverse momentum component while preserving its overall dispersion relation, as illustrated in Fig.~\ref{fig:boost}.

Explicitly, this relation reads
\begin{equation}\begin{aligned}
    \left(\begin{array}{l}
         E \\
         k_{\perp} 
    \end{array}\right) &=\left( 1 + \left(\begin{array}{cc}
        0  & \eta\\
         \eta & 0  
    \end{array}\right)\right) \left(\begin{array}{l}
         k_z \\
         0 
    \end{array}\right) =  exp\left[ \eta \left(\begin{array}{cc}
        0  & 1\\
         1 & 0  
    \end{array}\right)\right]\left(\begin{array}{l}
         k_z \\
         0 
    \end{array}\right) \\
    &= \left[\cosh \eta \left(\begin{array}{cc}
        1  & 0\\
         0 & 1  
    \end{array}\right) + \sinh\eta \left(\begin{array}{cc}
        0  & 1\\
         1 & 0  
    \end{array}\right)\right]\left(\begin{array}{l}
         k_z \\
         0 
    \end{array}\right) \\ 
    &= \left(\begin{array}{l}
         k_z\cosh \eta \\
         k_z \sinh \eta
    \end{array}\right).
    \end{aligned}
\end{equation}
Here, $\eta$ serves as a continuous boost parameter. In what follows, we discuss the transformation of the free-photon basis vectors to their counterparts inside the medium via boost-like operators. 

The Lorentz transformation operator is defined as\cite{Peskin1995AnIT}:
\begin{equation}
   S( \Lambda) = e^{-\frac{i}{2}\omega_{\mu \nu }S^{\mu \nu }},
\end{equation}
where $\omega_{\mu\nu}$ is an antisymmetric tensor representing the infinitesimal transformation parameters, and $S^{\mu \nu}$ are the corresponding generators. While the boost-like construction considered here formally resembles the Lorentz boost for Dirac spinors in electronic systems, a key distinction is that the corresponding generator is off-diagonal. It takes the form
\begin{equation}\label{boost}
    S^{\perp} = i\left(\begin{array}{cc}
        & \sigma_{\perp} \\ 
        -\sigma_{\perp} &
    \end{array}\right)=i\gamma_{\perp}.
\end{equation}
The spin operator for photons is defined as
\begin{equation}
    \begin{aligned} S^{ij}= \epsilon^{ijk}S_k,\quad S_k = \left(\begin{array}{cc}
        \sigma^k &  \\
         & - \sigma^k
    \end{array} \right),
         \end{aligned}
\end{equation}
where $\sigma^i$ are the Pauli matrices. Since we choose the propagation direction to be along the $z$-axis, the eigenvalue of the spin operator $S_z$ corresponds to the helicity:
\begin{equation}
    \begin{aligned}
 S_z \mathbf{e}_{s}(k)  = \left(\begin{array}{cc}
            \sigma_3  &  \\
             &  -\sigma_3 
        \end{array} \right) \mathbf{e}_{s}(k) = s \mathbf{e}_s(k).
    \end{aligned}
\end{equation}
where $s = \pm $, indicating that $\mathbf{e}_s(k)$ are the helicity eigenstates of the photon field. It is worth noting that the photon spin operator here plays a role formally analogous to the Lorentz boost generator in the electronic system. For free photons (massless case), the spin eigenstates correspond to the electronic spinor basis boosted to the limiting velocity $c$. In contrast, in a medium, photons effectively acquire a nonzero rest mass, and their spin eigenstates then correspond to boosted spinor bases under subluminal Lorentz transformations.

The basis vector in the medium under the boost operation \eqref{boost} can be expressed as
\begin{equation}
\begin{aligned}\label{boost e_+}
    \mathbf{e}_{\pm}(k) &=  S(\Lambda) \mathbf{e}_{\pm}(k_z)\\ 
    &=\sqrt{2\lambda k_z } \exp\left[ \eta_{\pm} \left(\begin{array}{cc}
        0 & \sigma_{\perp}\\
         -\sigma_\perp & 0
    \end{array}\right)\right]\mathbf{e}_{\pm}.
    \end{aligned}
\end{equation} 
Substituting the vacuum basis vector \eqref{jishiliang +} and the medium basis vector \eqref{Dirac:orthogonal relation} into the boost formula \eqref{boost e_+}, we find that the boost parameter $\eta_\pm$ and the medium parameters satisfy the relations
\begin{equation}\begin{aligned}
    \sqrt{2\lambda k_z}\cosh \eta_\pm &=N(\lambda+ Q_0), \\
   \sqrt{2\lambda k_z}\sinh \eta_{\pm} &= NQ_{\mp2},
    \end{aligned}
\end{equation}
where $N$ is the overall normalization factor. Making use of the hyperbolic identity
\begin{equation}
      (\cosh^2 \eta_\pm -  \sinh^2 \eta_\pm) =1 .
\end{equation}
We find
\begin{equation}
   2\lambda k_z = N^2[(\lambda+ Q_0)^2-Q^2_2]=N^2[ 2\lambda(\lambda+Q_0)].
\end{equation}
It follows that the normalization is given by
\begin{equation}
  N =k_z/ \sqrt{(\lambda + Q_0)k_z},  
\end{equation}
which is consistent with the coefficient appearing in the explicit expression for the basis vectors \eqref{Dirac:orthogonal relation}.  

In addition, the distinction between the positive- and negative-spin states is encoded in the parameter $Q_{\pm 2}$. In particular, one finds that
\begin{equation}
    \frac{Q_{+2}}{Q_{-2}} = \frac{\sinh \eta_{-2}}{\sinh \eta_{+2}} =e^{2i\phi},
\end{equation}
which characterizes the relative phase between the two spin components. This phase factor is directly analogous to the birefringence effect in anisotropic media: when light propagates through a birefringent material, orthogonal polarization components accumulate different phase shifts, leading to a relative phase factor such as $e^{i\Delta \phi}$. In our formalism, the factor $e^{2i\phi}$ plays a similar role — it encodes how the medium induces a relative phase shift between the two spin (helicity) states.

Repeating the same procedure for the conjugate basis vectors $\hat e_s(k_z)$ yields analogous results. Therefore, in close analogy with the electron case—where the spinor solutions of a stationary electron are boosted to finite-momentum spinors—we arrive at the compact relations for photons:
\begin{equation}
    \begin{aligned}
        \mathbf{e}_{s}(k) =S(\Lambda) \mathbf{e}_s(k_z),  \quad \mathbf{e}_s(k_z) =\sqrt{2\lambda k_z } \mathbf{e}_s;\\ 
        \widetilde{\mathbf{e}}_{s}(k) =S(\Lambda) \widetilde{\mathbf{e}}_s(k_z),  \quad \widetilde{\mathbf{e}}_s(k_z) =\sqrt{2\lambda k_z } \widetilde{\mathbf{e}}_s.
    \end{aligned}
\end{equation}

We also define the conjugate basis as $\overline{\mathbf{e}}( k) = \mathbf{e}^{\dagger}(k) \gamma^0$. The following orthogonality relations hold for the products:
\begin{equation}
    \begin{aligned}\label{orthogonality relation}
     \overline{\mathbf{e}}_s(k)\cdot \mathbf{e}_{s^{\prime}}(k ) = 2 \lambda k_z \delta_{ss^{\prime}},  \quad \mathbf{e}^{\dagger}_s(k) \mathbf{e}_{s^{\prime}}( k) =2 Q_0 k_z\delta_{ss^{\prime}}.
    \end{aligned}
\end{equation}
This structure is slightly different from that of electronic systems.
In the Dirac theory for electrons:
\begin{itemize}
    \item the quantity on the right-hand side of the first relation corresponds to the mass $m$,
    \item the quantity on the right-hand side of the second relation corresponds to the energy $E_{\bm k}$.
\end{itemize}
Here, however, these roles are interchanged: $2\lambda k_z$ plays the role of the energy, while $2 Q_0 k_z$ plays the role of the effective mass. Finally, summing over spin states and their conjugates yields the completeness relation:
\begin{equation}
    \begin{aligned}
        \sum_s \mathbf{e}_s(k) \overline{\mathbf{e}}_s(k) = \gamma^0 Q_0 k_z + \gamma_{\perp} Q_{\perp}k_z + \lambda k_z.
    \end{aligned}
\end{equation}

We also consider the case of negative-energy solutions, where $\omega = -\lambda k_z$. The corresponding solutions are
\begin{equation}
    \begin{aligned} \widetilde{\mathbf{e}}_-(k) = \frac{k_z}{\sqrt{(- \lambda + Q_0)k_z} }\left(\begin{array}{l}
        -\lambda + Q_0  \\
       \quad 0\\ 
        \quad 0\\ 
       \quad Q_{-2}    
    \end{array}\right), \quad \widetilde{\mathbf{e}}_+(k) = \frac{k_z}{\sqrt{( -\lambda + Q_0)k_z} }\left(\begin{array}{l}
        \quad 0 \\
        - \lambda + Q_0   \\ 
         \quad Q_{+2} \\ 
        \quad 0    
    \end{array} \right).
    \end{aligned}
\end{equation}
The orthonormality conditions for these negative-energy states are
\begin{equation}
    \begin{aligned}\label{anti orthogonality relation}
     \overline{ \widetilde{\mathbf{e}}}_s (k)\cdot \mathbf{e}_{s^{\prime}}( k ) = - 2 \lambda k_z \delta_{ss^{\prime}},\quad \widetilde{\mathbf{e}}^{\dagger}_s (k) \mathbf{e}_{s^{\prime}} ( k) =2 Q_0 k_z \delta_{ss^{\prime}}.
    \end{aligned}
\end{equation}
Summing over the spin states for the negative-energy solutions gives the completeness relation:
\begin{equation}
    \begin{aligned}
        \sum_s \widetilde{\mathbf{e}}_s(k) \overline{\widetilde{\mathbf{e}}}_s(k) = \gamma^0 Q_0 k_z + \gamma_{\perp} Q_{\perp}k_z - \lambda k_z.
    \end{aligned}
\end{equation}
For a detailed analogy between electrons and photons, see table~\ref{table}. In quantum electrodynamics (QED), the interaction between electrons and photons is described through the electron wavefunction $\psi_e$, the electric charge $e$, and the electromagnetic potential $\mathbf{A}^{el}$. In contrast, our work establishes a QED-like theoretical framework for photons themselves, formulated in terms of the photon wavefunction $\psi$, the helicity $s$, and medium-induced gauge field $\mathbf{A}^{me}$. Within this framework, the spin–orbit coupling of light can be reinterpreted as a gauge interaction analogous to that in QED~\cite{yang2025spinorbitcouplinghelicalwaveguides}.

\begin{table}[ht]\label{table}
\centering
\footnotesize
\begin{tabular}{lcc}
\toprule
\textbf{Category} & \textbf{Electronic System} & \textbf{Optical System} \\
\midrule

Lagrangian &
$\mathcal{L} = \overline{\psi} (i\gamma^{0}\partial_{0} + i\gamma^i \partial_i - m ) \psi$ &
$\mathcal{L} = \overline{\psi} (i\gamma^{0}\partial_{0} -\hat{m} - \gamma^0 \gamma^{\perp} \hat{p}_{\perp} ) \psi$ \\

Statistics &
Fermion: $\{ a_s,a^{\dagger}_{s^{\prime}} \}=\{ b_s,b^{\dagger}_{s^{\prime}} \}=\delta_{ss^{\prime}}$ &
Boson: $[a_s,a^{\dagger}_{s^{\prime}} ]=[ b_s,b^{\dagger}_{s^{\prime}} ]=\delta_{ss^{\prime}} $ \\

Energy relation &
Rest electron: $E_p^2 = m^2$ &
Free photon: $E_k^2 = \lambda^2 k_z^2$ ($\lambda=1$)\\

Energy relation &
Electron with momentum: $E_p^2 = m^2 + p^2$ &
Photon in a medium: $Q_0^2 k_z^2= \lambda^2 k_z^2+ Q_2^2 k_z^2$ \\[0.4em]

Hamiltonian &
$H=\psi^{\dagger}i\partial_t \psi$ &
$H=\psi^{\dagger}\gamma^0 i\partial_t \psi$ \\

Orthogonality &
$\mathbf{e}^{\dagger}_s \mathbf{e}_{s'} =2 E_p \delta_{ss'} ,\,$ $\overline{\mathbf{e}}_s\cdot \mathbf{e}_{s'} = 2 m \delta_{ss'}$ &
$\mathbf{e}^{\dagger}_s \mathbf{e}_{s'} =2 Q_0 k_z  \delta_{ss'},\,$ $\overline{\mathbf{e}}_s\cdot \mathbf{e}_{s'} = 2\lambda k_z \delta_{ss'}$\\

Interaction &
$\mathcal{L}_I\propto e\, \overline{\psi}_e \gamma^{\mu}A_{\mu} \psi_e$ &
$\mathcal{L}_I\propto \overline{\psi}\, s\, \gamma^{\mu}A^e_{\mu} \psi$ \\

\bottomrule
\end{tabular}
\caption{Analogy between the electronic system and the photonic system in the framework of the Dirac-like theory. The left column represents the electronic system, while the right column corresponds to the photonic system.}
\label{tab:compare}
\end{table}




\section{Quantization of Electromagnetic Observables in a Medium}
\subsection{Helicity, Energy, Momentum, and Orbital Angular Momentum}
It is worth noting that, similar to Dirac’s theory—where electric charge is a conserved quantity—helicity can likewise be associated with a conserved charge in a spin-degenerate medium without non-electromagnetic couplings\cite{FernandezCorbaton2013}. This conserved charge is defined as
\begin{equation}
    Q = \int d^3x J^0(x,t),
\end{equation}
where the conserved current related to helicity is given by $J^0 = s\psi^{\dagger} \gamma^0 \psi$(By replacing helicity $s$ with the electric charge $e$ and substituting the electron wave function accordingly, the corresponding conserved charge reduces to the conventional electric charge). The conservation of this current can be explicitly verified:
\begin{equation}
    \begin{aligned}\label{current}
        \partial_0 j^0 &= \partial_0 (\psi^{\dagger})\gamma^0 \psi + \psi^{\dagger} \gamma^0\partial_0 \psi \\
        &=i\psi^{\dagger} (\gamma^0  H^{\dagger} \gamma^0) \psi -i \psi^{\dagger} H \psi\\
        &=0,
    \end{aligned}
\end{equation}
where we have used the $\gamma^0$-pseudo-Hermitian condition (\ref{gamma0}).

We can use the mode expansion of the field to represent the conserved charge $Q$ as an operator:
\begin{equation}
    Q= \int d^3x \sum_s s (a^{\dagger}_sa_s - b^{\dagger}_sb_s).
\end{equation}
where the orthogonality relations (\ref{orthogonality relation}) and (\ref{anti orthogonality relation}) have been applied. This expression makes it clear that the contribution of antiphotons to the helicity charge carries the opposite sign from that of photons\cite{yang2025inducedberryconnectionphotonic}.
The total energy of the photon in the medium can be written as
\begin{equation}
    \begin{aligned}\label{QM: media energy}
        H = \int d^3r \psi^{\dagger} i \gamma^0 \frac{\partial \psi}{\partial t} = \int d^3p \sum_s E_{\mathbf{k} } \left( a^{\dagger}_s a_s + b^{\dagger}_s b_s \right ) + (\text{constant}) ,  
    \end{aligned}
\end{equation}
where the result follows from the orthogonality relations in Eqs.~(\ref{orthogonality relation}) and (\ref{anti orthogonality relation}).Unlike the case of a photon in free space, where the dispersion relation is simply $E = k_z$, the interaction with the medium modifies the energy to $E_{\bm{k}} = \lambda k_z$. This modified form closely parallels the relativistic energy–momentum relation for electrons, $E_p = \sqrt{m^2 + p^2}$, highlighting the effective “mass-like” contribution induced by the medium.

The momentum operator can be expressed as
\begin{equation}
    \begin{aligned}\label{QM: media momentum}
        \mathbf{P} =  \int d^3 r \psi^{*}(r)\gamma^0( -i\boldsymbol{\nabla} )  \psi(r) = \int d^3 k \sum_s  \mathbf{k} (a^{\dagger}_s a_s  +b^{\dagger}_s b_s),
    \end{aligned}
\end{equation}
where the orthogonality relations (\ref{orthogonality relation}) have been applied.
It is important to note that the momentum here is not simply given by the Poynting vector, $\mathbf{P} = \int d^3 r\mathbf{E} \times \mathbf{B}$, as discussed in Chapter~\ref{momentum}.
Within our wavefunction formalism, the electromagnetic momentum is directly related to the Minkowski momentum~\cite{PhysRevLett.AM,Pfeifer2009ConstrainingVO}, defined as $\bm{P}_{\mathrm{Min}} = \int d^3r \bm{D} \times \bm{B}$.
This quantity differs from the Abraham momentum, $\bm{P}_{\mathrm{Abr}} = \int d^3r\bm{E} \times \bm{H}$, and the relationship between the two can be understood through the presence of a medium-induced gauge field, which modifies the canonical versus kinetic structure of the electromagnetic momentum(in preparation).

When considering the influence of the gauge field induced by the geometric phase on photons, the expectation value of the total orbital angular momentum $\mathbf{J}$ can be written as \cite{BialynickiBirula1987BerrysPI}:
\begin{equation}
    \begin{aligned}
        \langle 1ph \vert \mathbf{J} \vert 1ph \rangle = \int d^3k ( f^*(k) (\mathbf{D_k}\times \mathbf{k}) f(k) +f^{\prime}(k)^*  (\mathbf{D_k}\times \mathbf{k}) f^{\prime}(k) ),
     \end{aligned}
\end{equation}
where the covariant derivative in momentum space is defined as
\begin{equation}
    \begin{aligned}
        \mathbf{D}_k = \frac{\partial  }{\partial \mathbf{k} }   + i \lambda  \bm{\alpha} ,
    \end{aligned}
\end{equation}
and the associated gauge field $\bm{\alpha}$ satisfies
\begin{equation}
    \begin{aligned}
        \mathbf{\nabla}_{\bf{k}} \times \bm{\alpha} = i s \frac{\mathbf{k}  }{k^3}.
    \end{aligned}
\end{equation}
These “momentum-space monopole” results are in agreement with those obtained from the perspective of noncommutative quantum mechanics. By introducing the position operator $        x_i \sim \frac{\partial }{\partial k^i} + i \lambda \alpha_i,$
one finds that the spatial coordinates acquire an intrinsic noncommutativity\cite{Skagerstam1992LocalizationOM,Berard:2004xn,Duval2001ExoticGS,Brard2003monopoleAB,Duval2004NoncommutingCE}. In this formulation, the coordinate algebra becomes\cite{Skagerstam1992LocalizationOM,Berard:2004xn,Duval2001ExoticGS,Brard2003monopoleAB,Duval2004NoncommutingCE}
\begin{equation}
    \begin{aligned}\label{noncommutativity}
    [x_i,x_j] =i\lambda\epsilon_{ijl} \frac{k_l}{k^3}.
    \end{aligned}
\end{equation}
which is precisely the commutation relation of a particle moving in the field of a monopole located in momentum space.

\subsection{Transverse Spin}

When constructing the spin operator in Eq.~\eqref{spin operator}, we restricted our analysis to plane waves whose spin points purely along the propagation direction. Here we extend the discussion to the transverse spin of evanescent waves by mapping evanescent waves onto plane waves defined within an effective medium. For example, consider an evanescent wave propagating along the $z-$ direction, while decaying in the $x-$ direction in vacuum, described by the field \cite{Bliokh2013ExtraordinaryMA}
\begin{equation}
    \begin{aligned}
        \mathbf{E}=\frac{A_0}{\sqrt{1+|m|^2}}\left(\overline{\mathbf{x}}+m \frac{k}{k_z} \overline{\mathbf{y}}-i \frac{\kappa}{k_z} \overline{\mathbf{z}}\right) \exp \left(i k_z z-\kappa x\right),
    \end{aligned}
\end{equation}
where $m$ denotes the polarization parameter. Since the electric field $\mathbf{E}$  rotates within both the transverse plane and the $x-z$ plane, the endpoint of the electric-field vector traces a cycloidal path. This generates a spin angular momentum component along the transverse $y-$axis, which is orthogonal to both the real and imaginary parts of the evanescent-field wavevector\cite{Bliokh_2015,Bliokh2013ExtraordinaryMA}.

This evanescent wave in vacuum can be modeled as a propagating plane wave in an effective medium characterized by parameters $Q_0$ and $Q_{\pm2}$. The dispersion relation follows as
\begin{equation}
 \lambda^2 k^2_z = k^2_z -\kappa^2 =k^2,
\end{equation}
with effective parameters 
\begin{equation}\label{Q_2}
    Q_0 =1, Q_2 =\pm \frac{\kappa}{k_z} .
\end{equation}
We consider either the purely antisymmetric solution $Q_{+2}=-Q_{-2}=iQ_{2y}$($\phi=\pi/2$) or the purely symmetric solution $Q_{+2}=Q_{-2}=Q_{2x}$ ($\phi=0$). The general case with both $Q_{2x}$ and $Q_{2y}$ are present, corresponds to a superposition of these two eigenstates, representing an eigenstate oriented along an arbitrary direction $\bm n$. These eigenstates in different orientations are related by the rotation operator$R(\theta)$, which transforms one eigenbasis into another.

The eigenvalues can be obtained from the dispersion relation:
\begin{equation}
    \lambda_{\pm} =\pm \frac{k}{k_z},
\end{equation}
where the positive and negative signs correspond to photons with positive and negative helicity, respectively. According to the definition of spin and the orthogonality conditions (see Eq.~\ref{orthogonality relation}), the components of the spin along the momentum direction are given by:
\begin{equation}\label{spin of evanescent wave}
    S_z = \sum_s(\lambda_+sa^{\dagger}_s a_s+\lambda_-sb^{\dagger}_sb_s ) .
\end{equation}
For a plane wave, one recovers $\lambda_{\pm } =\pm 1$. In the evanescent case, the replacement $\hat{\mathbf{k}} \rightarrow k/k_z$ leads to the average spin component\cite{Bliokh_2015}
\begin{equation}
    \langle S_z \rangle\propto s\frac{k}{k_z}.
\end{equation}

Next, we turn to the discussion of the transverse spin. The corresponding spin operator can be written as
\begin{equation}
    S^{ij}= \epsilon^{ijk} \left(\begin{array}{cc}
        \sigma^k &  \\
         & -\sigma^k
    \end{array}\right) = \epsilon^{ijk}S_k.
\end{equation}
Within this effective spin$-1/2$ description, the transverse spin of photons naturally emerges. This transverse angular momentum also contributes to the system’s dynamics, leading to additional effects in spin–orbit interaction \cite{yang2025inducedberryconnectionphotonic}. 


Specifically, the spin operator along the $x$-direction is given by
\begin{equation}
           S_x =\left( \begin{array}{cc}
            \sigma_x & 0 \\ 
            0 & -\sigma_x 
         \end{array} \right).
\end{equation}
The eigenstates of the spin along the $x$-axis can be written in a Pauli matrix-like form. In this case, they can be expressed in terms of the eigenstates of the spin along the $z$-axis as
\begin{equation}
    \mathbf{e}^+_x = \frac{1}{\sqrt{2}} (\mathbf{e}_+ + \mathbf{e}_-),\quad \mathbf{e}^-_x = \frac{1}{\sqrt{2}} (\mathbf{e}_+ -\mathbf{e}_-).
\end{equation}
Their explicit forms are given by
\begin{equation}
    \begin{aligned}\label{Sx}
 \mathbf{e}^+_x \propto \left(\begin{array}{l}
            \lambda + Q_0 \\ 
            \lambda + Q_0 \\ 
              - i Q_{2y} \\
              i Q_{2y}
         \end{array}\right), \quad \mathbf{e}^-_x\propto\left(\begin{array}{l}
            (\lambda + Q_0) \\ 
            -(\lambda + Q_0) \\ 
               i Q_{2y} \\
              i Q_{2y}
         \end{array}\right). 
    \end{aligned}
\end{equation}
 It is straightforward to verify that
\begin{equation}
    \begin{aligned}\label{S_x}
S_x \mathbf{e}^{\pm}_x = \pm \mathbf{e}^{\pm}_x .
    \end{aligned}
\end{equation}
Similarly, the spin operator along the $y$-direction is defined as
\begin{equation}
           S_y =\left( \begin{array}{cc}
            \sigma_y & 0 \\ 
            0 & -\sigma_y 
         \end{array} \right),
\end{equation}
with the corresponding eigenstates expanded in terms of the spin-$z$ eigenstates as
\begin{equation}
    \mathbf{e}^+_y = \frac{1}{\sqrt{2}} (\mathbf{e}_+ + i\mathbf{e}_-),\quad e^-_y = \frac{1}{\sqrt{2}} (\mathbf{e}_+ -i \mathbf{e}_-).
\end{equation}
Their explicit forms are given by
\begin{equation}
    \begin{aligned}\label{Sy}
\mathbf{e}^+_y  \propto\left(\begin{array}{l}
            \lambda + Q_0 \\ 
            i(\lambda + Q_0) \\ 
              - i Q_{2x} \\
              - Q_{2x}
         \end{array}\right), \quad \mathbf{e}^-_y\propto\left(\begin{array}{l}
            (\lambda + Q_0) \\ 
            -i(\lambda + Q_0) \\ 
              + i Q_{2x} \\
               -Q_{2x}
         \end{array}\right) .
    \end{aligned}
\end{equation}
A direct calculation then yields
\begin{equation}
    \begin{aligned}\label{S_y}
S_y \mathbf{e}^{\pm}_y = \pm \mathbf{e}^{\pm}_y.
    \end{aligned}
\end{equation}

\begin{figure}[htbp]
            \centering
            \includegraphics[height=0.28\textheight]{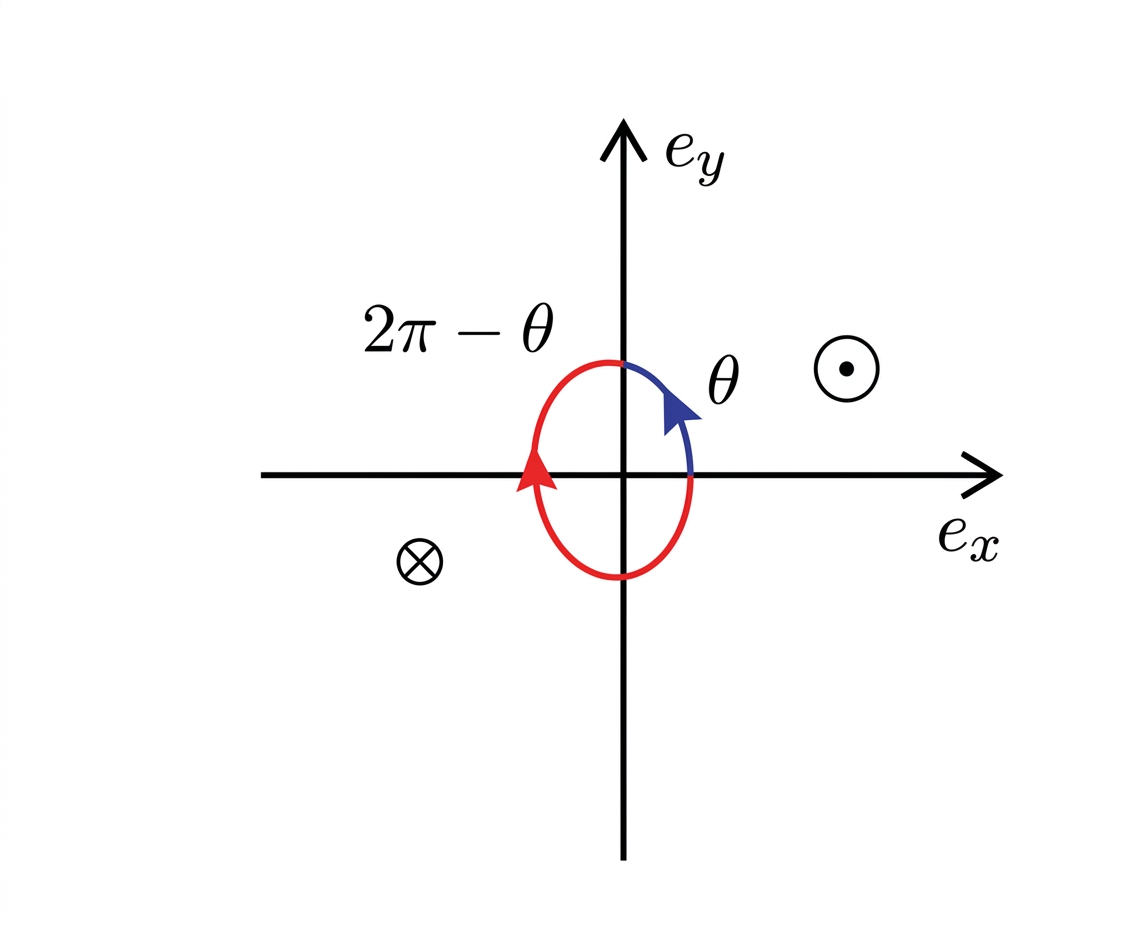}
             \caption{As illustrated, for a photon propagating out of the page along the direction denoted by $\bigodot$, a rotation through $\theta = \pi/2$ transforms the $x$-axis into the $y$-axis. In contrast, the antiphoton component propagates into the page ($\bigotimes$), and a rotation through $2\pi - \theta$ is needed to produce the identical $x\rightarrow y$ axis transformation. This correspondence holds for an arbitrary rotation angle $\theta$ and any direction $\bm n$.}
             \label{fig:R}
\end{figure}

Due to the attenuation along the $x$-direction, we choose the parameter $Q_{2x}$, which directly corresponds to the eigenstate with spin oriented along the $y$-direction. Naturally, if we perform a coordinate rotation that transforms the $x$-axis into the $y$-axis, this corresponds to attenuation along the $y$-direction. It can be verified that such a coordinate transformation precisely corresponds to rotating the basis vector from the $x$-direction to the $y$-direction, i.e., transforming a spin state originally aligned along $y$ into one aligned along $x$, with the effective parameter replaced by $Q_{2y}$. These two sets of basis vectors are therefore related by a rotation operator, which can be expressed as
\begin{equation}
    R(\theta) = \exp[-i\theta\frac{\bm J}{2} \cdot \bm n], 
\end{equation}
where $\bm J$ is the total angular momentum operator and $\bm n$ is the unit vector along the rotation axis.
However, an important point to note is that for the antiphoton component, its momentum — and thus its helicity — is opposite to that of the photon. As a result, when the photon part is rotated by an angle $\theta$ about the $z$-axis, the antiphoton part should instead be rotated by $2\pi - \theta$ see figure \ref{fig:R}. In particular, for a rotation angle $\theta = \pi/2$, we have
\begin{equation}
    R(\pi/2) = \left(\begin{array}{cccc}
       e^{-i\frac{\pi}{4}}  &  & &\\
         & e^{i\frac{\pi}{4}} & & \\
         & & e^{-i\frac{3\pi}{4}}& \\ 
         & & & e^{i\frac{3\pi}{4}}
    \end{array}\right).
\end{equation}
It can be explicitly verified that this rotation transforms the eigenstates from the $x$-direction basis into the $y$-direction basis, while satisfying the relation $Q_{2x}=iQ_{2y}$, up to an overall global phase factor:
\begin{equation}
    R(\pi/2) \mathbf{e}^{\pm}_x =e^{-i\frac{\pi}{4}}\mathbf{e}^{\pm}_y.
\end{equation}
Therefore, the effective parameter $Q_2$ represents the average transverse spin of the evanescent wave. This result is consistent with the known expressions for the transverse spin of light \cite{Bliokh_2015,Bliokh2013ExtraordinaryMA}.

\section{Conclusion}

Building on the Dirac-like formulation of Maxwell’s equations established in our previous work, we extend the framework by carrying out a second quantization of the photonic field, analogous to the procedure used in electronic systems. By introducing the single-photon state, we show that the expectation values of the energy, momentum, angular momentum, and spin operators of the photonic system in this state coincide with the corresponding physical quantities of the classical electromagnetic field. Moreover, this formulation provides a natural interpretation of the transverse spin of light, revealing its close analogy to the electron spin in quantum mechanics.

Furthermore, by drawing an analogy with the electronic field, we identified two remarkable correspondences. First, the helicity basis vector of a free photon shares the same mathematical form as the spinor solution of a free electron in its rest frame. In the electronic case, the spinor solution at arbitrary momentum is obtained through a Lorentz boost of the rest-frame solution. Similarly, we found that the helicity basis vectors of free photons can be transformed into those of photons in a medium by introducing an analogous "boost" operator. This structural similarity suggests that free photons and electrons at rest exhibit comparable mathematical forms at leading order. Moreover, in the presence of a medium, the dispersion relation of photons becomes analogous to that of accelerated electrons, indicating that photons effectively acquire a medium-induced mass.

Altogether, this framework provides a Dirac-like theoretical foundation for investigating light–matter interactions from a unified quantum perspective. At the scale of light–atom interactions, conventional treatments typically quantize only the atomic degrees of freedom while modeling the electromagnetic field as a classical wave. In contrast, the present formalism enables a fully quantum description in which both light and matter are treated on equal footing. This unified quantization opens the door to exploring interaction mechanisms and dynamical effects that are inaccessible within semiclassical approaches. Our results reveal previously unexplored properties of photonic states and provide new perspectives for understanding light in complex media, as well as for exploring fully quantum light--matter interactions.

\ack

This work was supported by the National Key R\&D Program of China (grant No. 2024YFE0109802), the National Natural Science Foundation of China (grants Nos. 12175320,12375084), the Natural Science Foundation of Guangdong Province, China (Grant No. 2022A1515010280).

\appendix

\section{Representation of Classical Electromagnetic Quantities in Terms of Creation and Annihilation Operators} \label{app3}
In this appendix, the electromagnetic field is formulated within a four-vector framework through the introduction of creation and annihilation operators. The equivalence between the field expressions for energy, momentum, angular momentum, and spin and the quantum mechanical expectation values of their corresponding operators is subsequently demonstrated.

Since the electric and magnetic fields, $\mathbf{E}(\mathbf{r})$ and  $\mathbf{B}(\mathbf{r})$, are real, they can be expressed in terms of the wave function and its mode expansion Eq.~(\ref{wavefunction}) and Eq.~(\ref{bohanshu zhankai}) as
\begin{equation}
    \begin{aligned}\label{dianchang}
        &\mathbf{E}(\mathbf{r}) =\frac{1}{2}(\psi + \psi^* )\\ 
        &=\int d^3k,\frac{N(k)}{2}\sum_s \left( a_s \mathbf{e}_s +  b_s \widetilde{\mathbf{e}}^*_s\right)e^{-i\omega t+i\bm k \cdot \mathbf{r} }  +\left(a^{\dagger}_s\mathbf{e}^*_s + b^\dagger_s \widetilde{\mathbf{e}}_s\right)e^{i\omega t -i\bm k\cdot \mathbf{r} },
    \end{aligned}
\end{equation}
and
\begin{equation}
    \begin{aligned}
        \mathbf{B}(\mathbf{r}) &= \frac{-i}{2}\hat{S}_z (\psi-\psi^*)\\ 
        &=\int d^3k \frac{N(k)}{2}\sum_s -is(a_s\mathbf{e}_s -b_s \widetilde{\mathbf{e}}^*_s )e^{-i\omega t+i\bm k \cdot \mathbf{r} } + is(a^{\dagger}_s\mathbf{e}^{*}_s -b^{\dagger}_s\widetilde{\mathbf{e}}_s )e^{i\omega t -i\bm k\cdot \mathbf{r} }.\\
    \end{aligned}
\end{equation}
Here, $a^{\dagger}_s$ and $a_s$, $b^{\dagger}_s$ and $b_s$ are the creation and annihilation operators for photons and antiphotons, respectively, satisfying the commutation relations given in Eq.~(\ref{commutation relation}). The definition of the spin projection operator $\hat{S}_z$ is provided in ~(\ref{S_z}) and the relation~(\ref{120}) has been used in the derivation.

In vacuum, the energy of the electromagnetic field can be expressed in terms of the transverse electric field $\mathbf{E}$ and the magnetic field $\mathbf{B}$ as
\begin{equation}
    \begin{aligned} \label{energy}
        H &=\int d^3r  \frac{1}{2}(\mathbf{E}^*(\mathbf{r}) \cdot \mathbf{E}(\mathbf{r})  + \mathbf{B}^*(\mathbf{r}) \cdot \mathbf{B}(\mathbf{r})) \\ 
        &=\frac{1}{4}\int d^3k \sum_sE_k [\left( a_s \mathbf{e}_s +  b_s \widetilde{\mathbf{e}}^*_s\right)+\left(a^{\dagger}_s\mathbf{e}^*_s + b^\dagger_s \widetilde{\mathbf{e}}_s\right)]^2+ [-is(a_s\mathbf{e}_s -b_s \widetilde{\mathbf{e}}^*_s ) + is(a^{\dagger}_s\mathbf{e}^{*}_s -b^{\dagger}_s\widetilde{\mathbf{e}}_s )]^2\\ 
        & =\frac{1}{2}\int d^3k\sum_s E_k (a^{\dagger}_sa_s +a_sa^{\dagger}_s + b_sb^{\dagger}_s + b^{\dagger}_s b_s)\\
         &= \int d^3 k \sum_s E_k  (a^{\dagger}_s a_s + b^{\dagger}_s b_s ) + (constant ).
    \end{aligned}
\end{equation}
In the second step, we have used the Parseval–Plancherel identity
\begin{equation}\label{Parseval–Plancherel identity}
    \int d^3r\, \mathbf{F}^*(\mathbf{r} ) \cdot \mathbf{G}(\mathbf{r}) =(2\pi)^{-3}\int d^3 k \,\mathbf{F}(\mathbf{k} ) \cdot \mathbf{G}(\mathbf{k}),
\end{equation}
and substitute the explicit form of $N(k)$. Making further use of the canonical commutation relations Eq.~\eqref{commutation relation} and the orthogonality condition for helicity eigenstates Eq.~\eqref{orthonormality e_s}, we recast the Hamiltonian into its normal-ordered expression.

Next, we express the Poynting vector as
\begin{equation}
    \begin{aligned}
        \mathbf{P} &=\int d^3r \mathbf{E}(\mathbf{r}) \times \mathbf{B}(\mathbf{r})\\ 
        &=\frac{1}{2}\int d^3k\, \mathbf{k} \sum_s[\left( a_s \mathbf{e}_s +  b_s \widetilde{\mathbf{e}}^*_s\right)+\left(a^{\dagger}_s\mathbf{e}^*_s + b^\dagger_s \widetilde{\mathbf{e}}_s\right)] \times [-is(a_s\mathbf{e}_s -b_s \widetilde{\mathbf{e}}^*_s ) + is(a^{\dagger}_s\mathbf{e}^{*}_s -b^{\dagger}_s\widetilde{\mathbf{e}}_s )] \\
        &=\frac{1}{2}\int d^3k\, \mathbf{k} \sum_s [isa_s a^{\dagger}_s \mathbf{e}_s\times \mathbf{e}^*_s -is a^{\dagger}_s a_s \mathbf{e}^*_s \times \mathbf{e}_s] + [isb^{\dagger}_sb_s \widetilde{\mathbf{e}}_s\times \widetilde{\mathbf{e}}^*_s - is b_sb^{\dagger}_s\widetilde{\mathbf{e}}^*_s \times \widetilde{\mathbf{e}}_s]\\
        &=\frac{1}{2}\int d^3k \, \mathbf{k} \sum_s(a_sa^{\dagger}_s + a^{\dagger}_s a_s + b^{\dagger}_sb_s + b_s b^{\dagger}_s)\\
        &=\int d^3 k \sum_s \mathbf{k}(a^{\dagger}_s a_s + b^{\dagger}_s b_s). 
    \end{aligned}
\end{equation}
In the second step, we used the identity Eq.~\eqref{Parseval–Plancherel identity} and the following relations between basis vectors:
\begin{equation}\label{cross product relation}
    \widetilde{e}_s\times \widetilde{e}^{*}_s = e_s \times e^*_s = -is e_{\mathbf{k} }.
\end{equation}
These relations indicate that the propagation direction of the antiphoton mode is opposite to that of the photon mode.

Similarly, the angular momentum of the electromagnetic field can be separated into orbital and spin parts:
\begin{equation} \mathbf{J} = \mathbf{L} + \mathbf{S}. \end{equation}
The first term corresponds to the orbital angular momentum, and the second term corresponds to the spin.
By substituting the wave-function form of the vector potential $\mathbf{A}(\mathbf{r},t)$ into the above expression, one finds that the orbital angular momentum takes the expected classical form$\mathbf{L} = \langle \mathbf{r}\rangle \times \mathbf{P}$, while the spin part reads\cite{cohentannoudji1989PhotonsAA,KELLER20051}
\begin{equation} \label{TAM}\int d^3r , \mathbf{r} \times (\mathbf{E} \times \mathbf{B}) = \int d^3r \sum_i E_i (\mathbf{r} \times \nabla) A_i + \int d^3r , \mathbf{E} \times \mathbf{A}. \end{equation}
Therefore, the first term in the above equation corresponds to orbital angular momentum, and the second term corresponds to spin. Since we use the transverse field here, the electric field and the vector potential satisfy the relation:
\begin{equation}
    -\frac{\partial \bm A}{ \partial t} = \bm E.
\end{equation}
Therefore, For monochrome the vector potential $\mathbf{A}(\mathbf{r},t)$, we expand it in accordance with Eq.~\eqref{dianchang} as
\begin{equation}
    \begin{aligned}\label{shineng A}
        \mathbf{A}(r,t)  =\int d^3k, N(k) \frac{-i}{2\omega}[(a_s \mathbf{e}_s + b_s \widetilde{\mathbf{e}}^*_s)e^{-i\omega t+i\bm k \cdot \mathbf{r} } -  ( a^{\dagger}_s\mathbf{e}^*_s + b^{\dagger}_s \widetilde{\mathbf{e}}_s )e^{i\omega t -i\bm k\cdot \mathbf{r} } ].\\
    \end{aligned}
\end{equation}
Substituting $\mathbf{A}$ into the total angular momentum (\ref{TAM}), the first part is
\begin{equation}
\begin{aligned}
\mathbf{L} =&\int d^3 r\sum_{i} \mathbf{E}^*_s  (\mathbf{r}\times \boldsymbol{\nabla}) A_s \\
&=\frac{1}{2}\int d^3k \sum_s [(a_s \mathbf{e}_s + b_s \widetilde{\mathbf{e}}^*_s ) +  ( a^{\dagger}_s\mathbf{e}^*_s + b^{\dagger}_s \widetilde{\mathbf{e}}_s) ](\mathbf{r}\times \mathbf{k}) [(a_s \mathbf{e}_s + b_s \widetilde{\mathbf{e}}^*_s ) +  ( a^{\dagger}_s\mathbf{e}^*_s + b^{\dagger}_s \widetilde{\mathbf{e}}_s  ) ]\\ 
 &=\frac{1}{2}\int d^3k \sum_s (\mathbf{r}\times \mathbf{k})(a_sa^{\dagger}_s + b^{\dagger}_s b_s +a^{\dagger}_sa_s +b_s b^{\dagger}_s)  \\
& =\int d^3k \sum_s(\mathbf{r}\times \mathbf{k}) (a_sa^{\dagger}_s + b_s b^{\dagger}_s  ).
 \end{aligned}
\end{equation}
In the second step, the identity Eq.~\eqref{Parseval–Plancherel identity} was used to transform the spatial integral into momentum space, and the monochromatic relation $E_k=\omega$ was applied. In the third step, the completeness relation \eqref{orthogonality relation} was employed. The final expression thus represents the orbital angular momentum operator in terms of the photon and antiphoton number operators.
Finaly, according to the commutation relations between creation and annihilation operators, the expression is written in normal ordering and the constant factor is omitted.
From Eq.~(\ref{QM:momentum}) it can be seen that the above expression is equivalent to the expectation value form of the angular momentum operator, and it corresponds to the classical angular momentum expression:
\begin{equation}
    \mathbf{L} =\langle \mathbf{r}\rangle \times \bf{P}.
\end{equation}
Next, we calculate the spin part. Substituting Eqs.~(\ref{dianchang}) and (\ref{shineng A}) into the second term of Eq.~(\ref{TAM}), we obtain
\begin{equation}
    \begin{aligned}\label{diancichang s}
        \mathbf{S} &= \int d^3 r  \,\mathbf{E}^* \times \mathbf{A}\\
        &=\frac{1}{2}\int d^3k \sum_s\left( a_s \mathbf{e}_s + b_s \widetilde{\mathbf{e}}^*_s \right)+[\left(a^{\dagger}_s\mathbf{e}^*_s + b^\dagger_s \widetilde{\mathbf{e}}_s\right)]\times [-i(a_s \mathbf{e}_s + b_s \widetilde{\mathbf{e}}^*_s) +i( a^{\dagger}_s\mathbf{e}^*_s + b^{\dagger}_s \widetilde{\mathbf{e}}_s ) ]/2\omega \\ 
        &= \frac{1}{2}\int d^3 k \sum_s s(a_sa^{\dagger}_s + a^{\dagger}_s a_s -b^{\dagger}_s b_s -b_sb^{\dagger}_s)\mathbf{e}_k \\ 
        & = \int d^3 k  \sum_s \left(sa^{\dagger}_sa_s -sb^{\dagger}_sb_s\right) \mathbf{e}_k.
    \end{aligned}
\end{equation}
In the second step, the identity ~\eqref{Parseval–Plancherel identity} and the cross-product relation ~\eqref{cross product relation} were used.
The final expression was rewritten in normal-ordered form by applying the commutation relations between the creation and annihilation operators of photons and antiphotons.

We now discuss from the perspective of operator expectation values. Since we have chosen the propagation direction, or in other words, we are using the transverse electromagnetic field, the spin operator is no longer the generator of SO(3). The corresponding spin operator in this four-component form only has the propagation direction and can be expressed as:
\begin{equation}
     \hat{\bm S}^{\prime} = \left(\begin{array}{cc}
        \sigma^3 &  \\
         &  \sigma^3
    \end{array} \right).
\end{equation}
The expectation value form of the spin operator $\bm{\hat{S}}$ is
\begin{equation}
    \mathbf{S} = \int d^3 r \,\psi^*(r)\gamma^0\bm{\hat{S}}^{\prime}\, \psi(r).
\end{equation}
Since in the helicity-space representation, the basis vectors are the eigenvectors of the spin operator, and photons and antiphotons have opposite contributions, we can redefine the form of the spin operator as $\hat{\bm S} = \gamma^0\hat{\bm S}^{\prime}$. Thus, we have
\begin{equation}\label{S_z}
    \hat{S}_z = \left(\begin{array}{cc}
        \sigma^3 &  \\
         & - \sigma^3
    \end{array} \right).
\end{equation}
The corresponding eigenvalue equations for the helicity basis vectors are
\begin{equation}
    \begin{aligned}\label{120}
        \hat{S}_z \mathbf{e}_s = s\mathbf{e}_s,\quad \hat{S}_z \,\widetilde{\mathbf{e}}_s = -s\widetilde{\mathbf{e}}_s.
    \end{aligned}
\end{equation}
According to the definition that the helicity operator is the projection of the spin along the propagation direction, $h = \bm S \cdot \hat{\bm k} $, it follows that the helicity of a photon is opposite to that of an antiphoton. The operator $\hat{S}_z$ represents the spin component along the propagation direction, and the quantity $s$ in the above expressions corresponds to the helicity eigenvalue. This means that for a single-photon helicity state, the contributions of the photon and the antiphoton are opposite. We express this in terms of creation and annihilation operators as
\begin{equation}
    \begin{aligned}\label{Spin2}
        \mathbf{S}= \int d^3k  \mathbf{\hat{k}} (a_+^{\dagger} a_+ - a_-^{\dagger}a_- - b_+^{\dagger}b_+ + b_-^{\dagger}b_-) ,
    \end{aligned}
\end{equation}
which is the same as the spin expression given by the electromagnetic field in Eq.~(\ref{diancichang s}). In the above expression, both positive-spin photons and negative-spin antiphotons contribute equally to the total helicity, and the same holds for the opposite case.The above discussion concerns the spin arising from the transverse field. When the electric field has a nonzero longitudinal component, the spin acquires transverse components, requiring inclusion of the $z$-component of the wavefunction. This aspect will be discussed in the main text.

\nocite{*}
\bibliographystyle{iopart-num}
\bibliography{quantization}

\end{document}